\documentclass[preprint2]{aastex}

\usepackage{epsf,epsfig}

\slugcomment{Accepted to ApJ}

\shorttitle{Likelihood distribution}

\shortauthors{Tucci et al.}

\begin{document}

\title{Likelihood distribution for models with cosmological constant
from COBE data}

\author{Marco Tucci\altaffilmark{1,2}, Angela Contaldo\altaffilmark{1}, Silvio Bonometto\altaffilmark{2,3}}

\altaffiltext{1}{Physics Department, Universit\`a degli Studi di
Milano, Milano (Italy)}

\altaffiltext{2}{I.N.F.N., Via Celoria 16, I20133 Milano (Italy)}

\altaffiltext{3}{Physics Department G. Occhialini, Universit\`a degli Studi di
Milano--Bicocca, Milano (Italy)}

\begin{abstract}

Using COBE--DMR 4--year data, we find a general expression
yielding the likelihood distribution in the 3--dimensional
parameter space spanned by the spectral index $n$, the spectral amplitude
$a_{10}$ and the false--vacuum density parameter $\Omega_\Lambda$.
Using such simple expression, the range of possible normalizations,
within a given likelihood interval from top--likelihood normalization,
is readily found, with fair approximation, for any model
with total density parameter $\Omega_o = 1$ and assigned $n$
and $\Omega_\Lambda$.

\end{abstract}


\section{Introduction}
The COBE--DMR data, so far, provide the only all--sky map for
the cosmic microwave background radiation (CMBR) (Smoot et al. 1992). 
Using such data, the angular temperature fluctuation 
spectrum $C_l$ can be inspected for $l$ values from 2 to 30. 
In principle, cosmological model parameters should be adjusted
so to provide the best possible fit to all $C_l$. Unfortunately,
however, the situation is more complicated than so. As a matter of fact,
model independent $C_l$ values and error bars were provided
by Hinshaw et al. (1996), Bunn \& White (1997), Tegmark \& Hamilton
(1997), assuming that $l(l+1)C_l$ takes single values, in suitable 
$l$ intervals. 
Tegmark (1997) and G\'orski (1997) also provided angular spectrum
data and error bars, separately for all $l$ values.
Attempting to use such $C_l$ to fix model parameters, as outlined
also by G\'orski (1997), is far from trivial.
In fact, two main difficulties arise. First of all,
each $C_l$ distribution, around its maximum likelihood value,
is not Gaussian, as is witnessed by its non--symmetric $1\, \sigma$
error bars; deviations from a Gaussian behaviour are even stronger
beyond 1$\, \sigma$.
Furthermore, the $C_l$ estimators are strongly correlated
with each other. Therefore, even 
obtaining the likelihood
distribution for each $C_l$, up to a few $\sigma's$ around
their maximum likelihood values,
a simultanous direct fit to different $C_l$'s bears no clear
statistical significance. 
Then, if precise likelihood comparisons within a 
given class of models are to be 
made, one 
has little alternative to fitting
directly model parameters to COBE--DMR pixel data. 
Such data are however public and can be found on the NASA site
(http://space.gsfc.nasa.gov/astro/cobe/).

In this work we focused on models with total density
parameter $\Omega_o = 1$, due to contributions
of baryons (density parameter $\Omega_b$), cold--dark--matter
(CDM; density parameter $\Omega_c$), and false vacuum
(density parameter $\Omega_\Lambda = \Lambda /8\pi G\rho_{cr}$).
Recent outputs of BOOMERanG and Maxima I experiment (de Bernardis et
al. 2000, Hanany et al. 2000)
confirm that $\Omega_o$ values substantially different from
unity are disfavoured. 
$\Omega_c$ and $\Omega_\Lambda$, together
with the primeval spectrum index $n$
and the primeval spectrum normalization,
are the parameters which
mostly shape the $C_l$ spectrum. On the contrary,
up to $l \simeq 30$,
the dark--matter (DM) composition and
the Hubble constant $H$ bear a limited impact only.
Attention was recently concentrated on models with $\Lambda \neq 0$,
mostly because data on SN Ia (Riess et al. 1998, Perlmutter et al. 1999)
seem to favour a negative deceleration parameter.

Previous analyses of the likelihood distribution
in the $a_{10}$--$n$ plane were performed by various authors. 
The parameter $a_{10} = \sqrt{C_{10}}\,T_0\,$ accounts for 
primeval spectrum normalization in a simple way,
as its best--fitting value
is nearly independent of $n$ in the 4--year COBE data.

Assuming a pure Sachs \& Wolfe
spectrum (Sachs \& Wolfe 1967, Bond \& Efstathiou 1987),
G\'orski et al. (1994), Bennet et al. (1994), Wright et
al. (1994), 
Bond (1995) and Tegmark \& Bunn (1995) analysed the 2--year COBE data.
Under the same assumption, G\'orski et al. (1996, G96 hereafter), 
Bennet et al. (1996),
Wright et al. (1996) and Hinshaw et al. (1996), analysed the 4--year
COBE data. Such spectrum can be considered a reasonable approximation
for pure CDM or mixed models with $\Omega_o = 1$, for low $l$'s.
On the contrary, open or $\Lambda$ models are expected to behave
differently (see, e.g., Stompor \& G\`orski 1994).
A first attempt to determine the normalisation of $\Lambda$ models and
an upper limit on the density parameter $\Omega_\Lambda$ using the
2--year COBE data was performed by Stompor, G\'orski \& Banday (1995a,
b) and Bunn \& Sugiyama (1995).
But the most extended results on the likelihood
of $\Lambda$--models, based on 4--year COBE data,
were obtained by Bunn \& White (1997, BW hereafter).

Let us briefly summarize BW results, which were also
partially included in the popular CMBFAST code
(Seljak \& Zaldarriaga 1996), to show
why a further effort is needed to complete them.
BW define a function $D(x)=l(l+1)C_l$, with $x=\log_{10}l$,
so that $D(1) = 110\, C_{10}$, and consider the parameters
$D'$ and $D''$ in the expansion
\begin{equation}
D(x)\simeq D_1\big[1+D'(x-1)+{D'' \over 2}(x-1)^2\big].
\end{equation}
Accordingly, $D'$ and $D''$ essentially account for the first
and second derivative of $D(x)$ at $x=1$ ($l = 10$).
BW provide, also through an analytic fitting expression, the
maximum--likelihood normalization $D_1$ for models defined
by the values of $D',~D''$ parameters and
the likelihood distribution in $D'$--$D''$ plane, for the
top likelihood value of $D_1$.
Among 2--parameter fittings of COBE--DMR data, the
BW approach is quite effective in treating
$\Lambda$--models, as fixing $D'$ and $D''$ corresponds to
giving $n$ and $\Lambda$ and {\it viceversa}. 

However, BW provide
only a partial infomation on the likelihood of models
whose amplitude is close to (but not coincident with)
the best--fit amplitude for given $n$.
In fact, let $\bar n$--$\bar a_{10}$ yield the top--likelihood model
and $n_\pm = \bar n$$\pm \Delta n$
limit the  1$\, \sigma$ error bar along the $n$ axis,
as given by BW. They complete the information on
likelihood distribution, by stating that 1$\, \sigma$ error bars
(on amplitude) roughly correspond to 7$\, \%$. 

In order to determine the likelihood of a model defined 
by $n$--$a_{10}$ values close to $\bar n$--$\bar a_{10}$,
one has to assume that, al least within 1$\, \sigma$, the
likelihood distribution is Gaussian. If so, one may determine
first the likelihood of the intermedate model given by
$n$--$\bar a_{10}$. Then, still assuming a Gaussian distribution
along the $a_{10}$ axis, one may determine the further
likelihood decrease when passing from $n$--$\bar a_{10}$ to
$n$--$a_{10}$. As a matter of fact, both distributions,
along the $n$ and the $a_{10}$ axes, are significantly
non--Gaussian, as is shown by the shape itself of the top--likelihood
curve. This already affects the above procedure within
1$\, \sigma$, but makes it seriously unadequate above 1$\, \sigma$.

It may be important, instead, to know how the likelihood is
distributed among models with different
normalizations. For instance, varying the normalization
of a model with given $n$ and $\Lambda$, we may predict
different galaxy cluster number densities. A model 
under--(over--)producing clusters, for maximum--likelihood
normalization, might be in agreement with observations
if a different normalization is taken. If this occurs
for spectral index and normalization
within 1$\, \sigma$ from the top--likelihood values
$\bar n$--$\bar a_{10}$,
rejecting such model because of cluster predictions is
illegitimate. But also knowing the range of normalizations
allowed, within 1--2--3$\, \sigma$'s, for a model with given
$\Lambda$ and $n$ may be relevant, as well as determining,
for a given $n$ and normalization, which is the range of
$\Lambda$ values allowed at the  1--2--3$\, \sigma$ level.
Similar arguments can be made for the capacity
of a model to predict a fair amount of high--$z$
objects
or to fit the spectral slope parameter $\Gamma$,
yielding the ratio between mass variances at 8 and 25 $h^{-1}$Mpc
(see, e.g., Bonometto \& Pierpaoli 1998, for more details),
while other parameter combinations may also have to be explored.

Quite in general, to extend BW results in this way, we must perform
a 3--parameter analysis of COBE--DMR data. Our parameters are
the spectral index $n$, the normalization $a_{10}$, and the vacuum
density parameter $\Omega_\Lambda$. The result of our analysis will
be a handable analytic expressions of the likelihood distribution
in such 3--parameter space. We shall also provide
a simple numerical routine, which can be appended to CMBFAST
and gives, besides of the best--fit normalization
for fixed $n$ and $\Lambda$, also the limits of the
fluctuation amplitude intervals where likelihood decrements
corresponding to $n$ $\sigma$'s ($n=1,2,3$) occur.
Other details provided by such routine will be discussed below.

\section{From COBE data to the likelihood distribution}

In this section we shall report how the likelihood distribution
is obtained from COBE--DMR data. A large part of this section is
based on G\'orski (1994, G94 hereafter). Its details are unessential
to read the following sections and to use our relations yielding
the likelihood distribution. Let us rather draw the reader's
attention on the Appendix B, where we show in detail how the likelihood
dependence on monopole and dipole terms are integrated out.

COBE--DMR temperature data were estimated at the frequencies
of 31.5, 53, and 90 GHz. Using such estimates, the vectors ${\bf d}$,
with $N_p$ (pixel number) components $d_i$, can be built.
The signal variance, due to uncertainties of instrumental
origin at each pixel, $\sigma_i$, can also be collected in
$N_p$--dimensional {\it noise} vectors ({\boldmath $\sigma$}).
Each pixel is centered on a point of the celestial sphere
of coordinates $\hat n$$_i \equiv (\theta_i,\phi_i)$.
The likelihood of a given model $\cal M$ is obtained by comparing
${\bf d}$ with a {\it fictitious data} vector {\boldmath $\delta$}
(components $\delta_i$), built from the model, but taking into
account how real data are obtained and, therefore, also the noise
vector {\boldmath $\sigma$}. 
Noise correlation among pixels, expected to be small, will be neglected
in our theoretical developments. 

In principle this can be done by evaluating the matrix
${\bf M}$, whose components
\begin{equation}
M_{ij} = \langle \delta_i \delta_j \rangle 
\end{equation}
are obtained by averaging on the ensemble of $\cal M$ realizations.
Assuming a Gaussian statistics, the likelihood of $\cal M$ reads then:
\begin{equation}
{\cal L} = [(2\pi)^{N_p} det{\bf M}]^{-1/2}
\exp[-(1/2){\bf d}^{\rm T}{\bf M}^{-1}{\bf d}] ~.
\label{l1}
\end{equation}
A slight non--Gaussian behaviour in data (Ferreira, Maguejo \& G\'orski 1998,
Novikov, Feldman \& Shandarin 1998, Pando, Valls--Gabaud \& Fang 1998)
possibly originated by post--recombination processes, should not
affect our analysis, as shown in detail by Contaldi et al. (1999),
for a standard CDM model. 

However, eq.~(\ref{l1}) can hardly be used, because of the
high dimensionality of the matrix ${\bf M}$, which should be
inverted. Replacing the pixel basis (a discretized coordinate
representation) by an angular harmonic basis (essentially its
Fourier transform) allows to reduce the dimensionality of
matrices, without wasting physical information. When doing so,
we must also take into account that temperature fluctuation data 
about the galactic equatorial plane do not give cosmological 
information.

In order to evaluate the likelihood of a model $\cal M$, therefore,
the first step amounts to defining the set of real spherical harmonics:
\begin{equation}
Y_{lm} ({\hat r}) = \sqrt{l+1/2} \sqrt{{(l-|m|!) \over (l+|m|!)}} 
P_l^{|m|} (\cos \theta) f_m (\phi)
\end{equation}
where
\begin{equation}
f_m (\phi) = (2\pi)^{-1/2}\,,\pi^{-1/2} \cos(m\phi)
\,,\pi^{-1/2} \sin(|m|\phi)
\end{equation}
for $m=0,>0,<0$, respectively, while the
\begin{equation}
P_l^m (x) = (-1)^m (1-x^2)^{m/2} (d/dx)^{|m|} P_l(x)
\end{equation}
are obtained by differentiating the ordinary Legendre polynomials
$P_l$. Such spherical harmonics are built so to fulfill the
orthonormality conditions
\begin{equation}
\int_{4\pi} d^2 {\hat r} Y_{lm} (\hat r) 
Y_{l'm'} (\hat r) = \delta_{ll'} \delta_{mm'} 
\end{equation}
when the integration is extended on the whole sky.
If this integration is replaced by a sum on the $N_p$ pixel
centers, orthonormality is recovered by replacing $Y_{lm}$ by
\begin{equation}
{\hat Y}_{lm} ({\hat r}) = w_l^{pix} Y_{lm} ({\hat r})
\end{equation}
(a table of $w_l^{pix}$ and directions on their computation are
given in Appendix A).

Using such spherical harmonics, then, we
can build the components of a fictitious signal vector,
for the point $\hat r$$_i$, yielding the $i$--th pixel, as follows:
\begin{equation}
\delta({\hat r}_i) =  \sum_{l=0}^\infty \sum_{m=-l}^l 
(a_{lm}^{CMB} w_l^{DMR} + a_{lm}^{\rm noise}) {\hat Y}_{lm}({\hat r}_i) ~.
\label{y1}
\end{equation}
Here, besides of $a_{lm}^{CMB}$, that are to be obtained for
a realization of $\cal M$,
we need the $a_{lm}^{\rm noise}$ originated by the instrumental
noise and the window function $w_l^{DMR}$ for the observational apparatus
(given, e.g., by Wright et al. 1994). Accordingly, as above outlined,
the fictitious signal vector that we shall build will convolve the
features of a given cosmology with the characteristics of the
COBE--DMR apparatus.

It is also convenient to put together the indices $l$ and $m$, by defining
\begin{equation}
\mu = l(l+1) + (m+1) ~.
\end{equation}
If $l$ values up to $l_{max} = 30$ are considered, $\mu $ attains
the maximum value $\mu_{max} = (l_{max} + 1)^2 = 961$. Accordingly,
the functions ${\hat Y}_\mu$ define the passage from the
$N_p$--dimensional pixel basis to a 961--dimensional basis.
In the sequel, latin indices shall be used to indicate vector
components on the pixel basis only; accordingly, eq.~(\ref{y1}) can be
unambiguously rewritten as
\begin{equation}
\delta_i =  \sum_\mu ({\tilde a}_\mu^{CMB} + a_\mu^{\rm noise})
{\hat Y}_{\mu i} 
\label{y11}
\end{equation}
($\sim$ is used to indicate that window function effects
are already taken into account).

Let us then remind that the $a_{lm}^{CMB} \equiv a_\mu^{CMB}$, that we
shall build, depend on the model $\cal M$
and on its realization; ensemble averaging them, one would obtain the
angular spectrum components $C_l \equiv \langle |a_{lm}^{CMB}|^2
\rangle.$

The use of spherical harmonics, however, gives place to a problem.
When using real data, pixels within 20$^o$ from the galactic plane
are to be excluded (in the sequel, residual pixels will be
said to belong to the {\it cut--sky}). First of all, this
reduces $N_p$ from 6144 to 4016. Furthermore, in the cut sky,
the set of polynomials ${\hat Y}_{\mu}$ is no longer orthonormal. 
Orthonormality, however, can be recovered by replicing the basis
$\bf{\hat Y}$$\equiv \{\hat Y_\mu \}$ by the basis
$$
{\bf \Psi} = {\bf \Gamma} \cdot {\hat {\bf Y}}
$$
where ${\bf \Gamma}$ is a 961$\times$961 matrix 
that we shall now define.
In order to have $\langle {\bf \Psi} {\bf \Psi}^{\rm T}
\rangle_{c.s.} = {\bf I}$ ($\langle .... \rangle_{c.s.}$ is a
product obtained summing on the set of pixels belonging to the
cut--sky; ${\bf I}$ is the unit matrix), it must be
\begin{equation}
{\bf \Gamma} \cdot \langle {\hat Y}{\hat Y}^{\rm T} \rangle_{c.s.}
\cdot {\bf \Gamma}^{\rm T} = {\bf I}
\end{equation}
and this is obtained if the matrix $\langle {\hat Y}{\hat Y}^{\rm T}
\rangle_{c.s.}$ is expressed as a product ${\bf L} \cdot {\bf L}^{\rm
T}$ and 
\begin{equation}
{\bf \Gamma} = {\bf L}^{-1} ~.
\end{equation}
Here we shall use the upper--triangular matrix ${\bf \Gamma}$,
obtained by performing the so--called Kolewski decomposition, as
done by G94. This choice is not unique; alternative
possibilities are discussed by Tegmark (1997). However, as we shall
see, this choice allows a simple integration of the likelihood
on monopole and dipole data.

The fictitious data vector {\boldmath $\delta$}, can be expanded on both
${\bf Y}$ and ${\bf \Psi}$ bases; let $\tilde a$$_\mu$ and $c_\mu$
be its components, respectively. Clearly:
\begin{equation}
{\tilde a}_\mu = \sum_\nu \Gamma_{\mu \nu} c_\nu
~~{\rm and}~~
c_\mu = \sum_\nu L_{\mu \nu}  {\tilde a}_\nu ~.
\label{ac}
\end{equation}
Therefore, once Kolewski decomposition provides the ${\bf L}$ matrix,
we readily obtain $c_\mu
=\langle {\bf d}{\bf \Psi}_{\mu}\rangle_{c.s.}$; in particular,
besides of the CMB component $c_\mu^{CMB}$, the noise component
$c_\mu^{\rm noise}$ is soon obtainable. Let us also outline that,
thanks to the triangular form of the ${\bf L}$ matrix, the $c_\mu$
coefficients are linear combinations of $a_\nu$ coefficients with
$\nu \geq \mu$. 

Using the basis $\Psi_\mu$, eq.~(\ref{y11}) reads:
\begin{equation}
\delta_i = \sum_\mu c_\mu \Psi_{\mu i} =  \sum_\mu (c_\mu^{CMB} +
c_\mu^{\rm noise}) 
\Psi_{\mu i}  ~
\label{psi1}
\end{equation}
and
\begin{equation}
M_{\mu\nu} = \langle c_\mu c_\nu \rangle = 
\langle c_\mu^{CMB} c_\nu^{CMB} \rangle +
\langle c_\mu^{\rm noise} c_\nu^{\rm noise} \rangle 
$$$$
= M_{\mu\nu}^{CMB} + M_{\mu\nu}^{\rm noise}
\end{equation}
is the new correlation matrix (assuming that signal and noise are
both distributed in a Gaussian way and uncorrelated). Using 
$ M_{\mu\nu}^{CMB}$ and $ M_{\mu\nu}^{\rm noise}$, the
expression for the likelihood of the model reads:
\begin{eqnarray}
\label{l2}
{\cal L} = [(2\pi)^{N} det({\bf M}^{CMB} + {\bf
M}^{noise})]^{-1/2}\times \nonumber \\
~~\times\exp[-(1/2){\bf c}^{\rm T}({\bf M}^{CMB} + {\bf
M}^{noise})^{-1}{\bf c}] \,. 
\end{eqnarray}
Here ${\bf c}$ is a vector with 961 components $c_\mu$. 
Thanks to the triangular form of ${\bf L}$, we can easily
integrate the likelihood results on $c_\mu$ for $\mu = 1,..,4$
(such integration can be made in an exact way; details are given
in Appendix B). Hence, in the cut sky, the effective basis
dimensionality will be 957 (for each signal frequency
to be considered).

To build the likelihood eq.~(\ref{l2}),
$\tilde C_l$ are obtained using the CMBFAST program, convolved with
the COBE--DMR window function. Then, according to eq.~(\ref{ac}), 
\begin{equation}
{\bf M}^{CMB} = {\bf L}^{\rm T} \cdot \langle {\tilde {\bf a}}^{CMB}
{{\tilde {\bf a}}^{CMB~{\rm T}}} \rangle \cdot {\bf L}
= {\bf L}^{\rm T} \cdot {\bf C}^{CMB} \cdot {\bf L};
\end{equation}
here ${\bf C}^{CMB}$ is a diagonal matrix, whose diagonal terms
are the components of the angular spectrum  $\tilde C_l$,
however is repeated $2l+1$ times (for all $\mu$ values which
correspond to a given $l$). 
${\bf M}^{CMB}$ is then to be set together with the noise matrix
\begin{equation}
{\bf M}^{\rm noise} =
\langle {\bf c}^{\rm noise} {{\bf c}^{\rm noise ~T}} \rangle
= \Omega_{pix}^2 \sigma_i^2  {\bf \Psi}_{i}  {\bf \Psi}_{i}^{\rm T}
\end{equation}
(the index $_i$ runs on the 4016 pixels of the cut--sky).

In this way, the likelihood of a given model $\cal M$ is finally
evaluated and different models may be tested against COBE--DMR data.

\section{Results}

{\begin{table*}
\centering
\begin{tabular}{ccccccccc}
\tableline
$h$ & $\Omega_{\Lambda}$ & $P_1$ &$P_2$ &$P_3$ &$P_4$ & $P_5$ &
$n^o$ & $a_{10}^o/\mu K$ \\
\tableline
\tableline
0.7 & 0.8 & 9.93 & 1.14 & 2.24 & -1.53 & 1.72 & 1.1711 & 6.1579 \\
   & 0.7 & 12.31 & 0.50 & 2.36 & -0.71 & 1.92 & 1.1149 & 6.2173 \\
   & 0.6 & 14.68 & -0.123 & 2.17 & 0.66 & 2.07 & 1.0831 & 6.2547 \\
   & 0.4 & 17.46 & -1.014 & 1.38 & 3.28 & 2.28 & 1.0550 & 6.2927 \\
   & 0.2 & 18.62 & -1.44 & 0.89 & 4.59 & 2.37 & 1.0489 & 6.3073 \\
\tableline
0.6 & 0.7 & 12.43 & 0.42 & 2.33 & -0.61 & 1.92 & 1.1138 &6.2176 \\
   & 0.6 & 14.69 & -0.168 & 2.14 & 0.73 & 2.07 & 1.0852 &6.2552 \\ 
   & 0.4 & 14.69 & -1.04 & 1.36 & 3.22 & 2.28 & 1.0597 &6.2944 \\ 
   & 0.2 & 14.69 & -1.46 & 0.88 & 4.49 & 2.36 & 1.0550 &6.3099 \\ 
\tableline
0.5 & 0.4 & 17.41 & -1.052 & 1.36 & 3.27 & 2.28 & 1.0605 & 6.2941 \\
   & 0.3 & 18.08 & -1.300 & 1.06 & 4.02 & 2.33 & 1.0574 & 6.3039 \\ 
   & 0.2 & 18.45 & -1.458 & 0.89 & 4.47 & 2.36 & 1.0574 & 6.3102 \\
   & 0.1 & 18.61 & -1.540 & 0.77 & 4.69 & 2.38 & 1.0598 & 6.3136 \\
   & 0.0 & 18.33 & -1.58 & 0.7 & 4.5 & 2.38 & 1.0647 & 6.3138 \\ 
\tableline
\end{tabular}
\caption{ Values of parameter to fit the likelihood function, see
eq. (\ref{piu}). Such values are reported as an
intermediate step of our work and mostly to show the residual
small dependence on $h$ of the fitting costants. Such dependence
will be neglected in the final fitting expression. \label{param}}
\end{table*}}

Using the approach described in the previous section, 
substantially coincident with G94 treatment,
we considered a lattice of models for $h$ varying from 0.5 to 0.7
and $\Omega_\Lambda$ from 0 to 0.8. For each
model, the likelihood was evaluated for different values of
$n$ and $a_{10}$. We used 4--year COBE--DMR data, performing
a weighted average of 53 and 90 GHz outputs; we verified that
using the noisy data at 31.5 GHz does not add any substantial
improvement. More precisely, in accordance with Tegmark \& 
Bunn (1995) and Bunn \& White (1997), we may set
\begin{equation}
d_i = {{d_i(53) \sigma_i^{-2}(53)} + {d_i(90)\sigma_i^{-2}(90)}
\over \sigma_i^{-2}(53) + \sigma_i^{-2}(90)}
\end{equation}
and
\begin{equation}
\sigma_i^{-2} = 
\sigma_i^{-2}(53) + \sigma_i^{-2}(90)
\end{equation}
as data and noise vector components, respectively.
Using such inputs, we worked out the
3--dimensional curves $\cal L$$(n,a_{10})$ for each model.
As an example, in Fig. \ref{3dlike} we show it for a model with
$h = 0.5$ and $\Lambda = 0$. From such curve, the isoprobability
contours can be readily obtained. They encompass
volumes corresponding to 68.3$\, \%$, 95.4$\, \%$, 99.7$\, \%$
of the total volume below the 2--dimensional curve, respectively,
and will be called 1--2--3$\, \sigma$ contours, in the sequel.

We found a fairly regular behaviour of $\ln($$\cal L$) which
can be expressed through a third degree polynomial in
$x\,$$\equiv n-n^o$ and $y\,$$\equiv a_{10} - a_{10}^o$;
here $n^o$ and $a_{10}^o$ yield the peak position.
More in detail, our fitting formula reads:
\begin{eqnarray}
\label{piu}
-2 \ln {\cal L}' = P_1 x^2 + P_2 xy + 5.9 y^2 + P_3 x^2 y + \nonumber \\
+ P_4 y^3 + P_5 xy^2 - 0.78 y^3 ~.
\end{eqnarray}
In Table \ref{param} we give the values of the best--fit values of
the parameters $P_i$ ($i=1,..,5$),
$n^o$ and $a_{10}^o$ for the models of the lattice.
The expression (\ref{piu}), together with the values in Table \ref{param},
outlines that substantial deviation from a Gaussian behaviour
in the error distribution are present; however, to put them
under control, up to the 3$\, \sigma$ level, it is sufficient to use
third degree polinomials. The quality of the fits
obtained in this way can be appreciated from Fig. \ref{likeh}, where
we show how the 1--2--3$\, \sigma$ contours obtained
from model analysis and fitting formula agree, for
a set of typical cases. Residual discrepancies, visible
for the 3$\, \sigma$ contours, correspond to overall
likelihood shifts $\sim 10^{-6}$. 

{\begin{table}
\begin{tabular}{llll}
\tableline
        & $n$ & $Q(\mu K)$ & $Q|_{n=1}(\mu K)$ \\
\tableline
\tableline
$G20\,*$ & $1.22^{+0.24}_{-0.28}$ & $16.71^{+3.93}_{-3.12}$ &
        $19.40^{+1.29}_{-1.25}$ \\ 
$G20+\,*$ & $1.23^{+0.23}_{-0.29}$ & $15.26^{+3.93}_{-2.64}$ &
        $18.34^{+1.25}_{-1.20}$ \\
$G30+\,*$ & $1.23^{+0.29}_{-0.34}$ & $15.18^{+4.46}_{-2.92}$ &
        $17.82^{+1.44}_{-1.34}$ \\
$G20+~l-g$ & $1.11^{+0.38}_{-0.42}$ & $16.33^{+5.18}_{-3.69}$ &
        $17.38^{+1.77}_{-1.68}$ \\
$G30+~l-g$ & $0.79^{+0.48}_{-0.55}$ & $18.63^{+7.72}_{-5.08}$ &
        $16.57^{+1.92}_{-1.82}$ \\
$G20+\,*\,-g$ & $1.21^{+0.24}_{-0.28}$ & $15.23^{+3.69}_{-2.64}$ &
        $17.67^{+1.25}_{-1.15}$ \\
$G30+\,*\,-g$ & $1.24^{+0.27}_{-0.33}$ & $14.80^{+4.07}_{-2.83}$ &
        $17.34^{+1.39}_{-1.34}$ \\
H $w$  & $1.25^{+0.26}_{-0.29}$ & $15.4^{+3.9}_{-2.9}$ &
$18.4^{+1.4}_{-1.3}$ \\ 
H $w\,-g$   & $1.23^{+0.26}_{-0.27}$ & $15.2^{+3.6}_{-2.8}$ &
        $17.8^{+1.3}_{-1.3}$ \\ 
H $l$ & $1.00^{+0.40}_{-0.43}$ & $17.2^{+5.6}_{-4.0}$ & $17.2^{+1.9}_{-1.7}$ \\
B $w2$ & $1.18\pm 0.28$ & $16.2$ & $18.7\pm 1.26$  \\
T20 $w2$  & $1.21^{+0.35}_{-0.42}$  & $14.7^{+5.56}_{-3.36}$  &
        $17.1\pm 1.5$ \\
\tableline
\end{tabular}
\caption{Best fit values of $n$ and $Q$ (quadrupole) from this
and previous works; the best--fit values of $Q$ for $n = 1$
are also given. In the first column we give the initial of the first
author (G for G96, H for Hinshaw et al. 1996, B for BW and T for the
present work). The number after G indicates the
width of the cut around the galactic equator and, in general, notation
is the same as in G96. Results by other authors are obtained by
operating the so--called customary cut
suggested by Bennet et al. (1996). The letter $w$
states that results were obtained through a weighted average
of the 6 maps (3 frequencies, 2 channels); $w2$ indicates
that only 4 maps were used, neglecting 31.5 GHz outputs;
$l$ indicates that a linear combination of results at
various frequencies was done, taking coefficients able to
cancel the contribute from free--free galactic emission;
$-g$ indicates that a correction for foregrounds was done;
results obtained considering separately results at 3
frequencies are marked with * (961+961+961 component
vectors). \label{respre}}
\end{table}}

The fact that the expected non--Gaussian behaviour is
so simply fitted, is not the only finding of this work. In fact,
we also find that all the coefficients $P_i$, $n^o$ and $a_{10}^o$ in
the expression (\ref{piu}), can be fairly approximated using a single
interpolating expression, as simple as
\begin{equation}
 a + b \Omega_\Lambda^S + c \Omega_\Lambda^{2S} ~.
\label{int}
\end{equation}
In Table \ref{param2} we give the values of the interpolating
coefficients $a$, $b$, $c$ and $S$. In Fig.~\ref{p2p4}, we show how
our fitting formula meets the $P_i$ values (in two typical cases),
$n^o$ and $a_{10}^o$ obtained for the various models.

{\begin{table}
\label{param2}
\begin{tabular}{cccccc}
\tableline
  & & $a$ & $b$ & $c$ & $S$ \\
\tableline
\tableline
 $P_0$ & & 8.34 & 0 & -8.4 & 1.89 \\
\tableline
 $P_1$ & & 18.35 & 4.1 & -19.5 & 1.14 \\
\tableline
 $P_2$ & & -1.574 & 4.58  & 0 &  2.30 \\
\tableline
 $P_3$ & & 0.74 & 7.4 & -8.4 & 2.54  \\
\tableline
 $P_4$ & & 4.66 & -22.3 & 20 & 2.96 \\
\tableline
 $P_5$ & & 2.383 & -1.19 & 0 & 2.64 \\
\tableline
 $n^o$ & & 1.06 & -0.09 & 0.38 & 1.9 \\
\tableline
 $a_{10}^o/\mu K$ & & 6.312 & -0.1 & -0.2 & 2 \\
\tableline
\end{tabular}
\caption{Best--fitting values of parameters in eq. (\ref{int})}
\end{table}}

In order to validate our algorithm, we produced 
120 CMBR sky realizations for 
cosmological models with assigned
$a_{10}=6.93 \, \mu K$, $\Omega_\Lambda = 0$ and $n = 1.00$,
simulated their observation with COBE--DMR, and applied
our algorithm to determine the model likelihood in the
space spanned by $a_{10}$, $\Omega_\Lambda$ and $n$.
The simulated CMBR sky was produced using an algorithm based
on the technique suggested by Muciaccia, Natali \& Vittorio (1997).
The $C_l$ coefficients required were generated using CMBFAST.
The spectrum was then multiplied by 
the COBE--DMR experiment window function.
COBE--DMR features were used also to define pixels and
pixel noise variance. Using such fictitious data, we
searched the top--likelihood point for all models
in the $n$--$a_{10}$ plane, using the
same algorithm applied to COBE--DMR data.
The effect of the 20$^o$ subtraction along the galactic 
plane was also tested, by searching the
top--likelihood point for all models, also without 
galactic plane subtraction.
Our algorithm recovers $n = 1.01 \pm 0.13$ and $a_{10}
= 6.92\, \mu K \pm 0.30\, \mu K$ (the latter standard deviation corresponds
to $\simeq 4\, \%$). When the full sky is used, standard
deviations are reduced by $\sim 20\, \%$ and $\sim 30\, \%$,
respectively.
For the sake of comparison, let us report that a more limited test,
reported by BW, who made simulated maps with purely Sachs \& Wolfe
input spectra, gave $\Delta n \sim 0.26$ and a normalization
discrepancy up to $\sim 7\, \%$.

We also compared our results
with those cases already treated in the literature. 
In Table \ref{respre} we present the results of previous works on $n$
and $Q$, using 4--year COBE--DMR data. The last line of the table
provides our estimate, assuming a pure Sachs \& Wolfe spectrum. 
The dispersion in the values of the table is to be attributed 
to different choices of the ``Galactic--plane cut'' and 
different combinations of the 3 frequencies, while results
should not be affected by different data compression techniques. 
For each result, however, we indicate the method used by the authors.

Here we shall report some details
on the comparison with the analysis of G96, on 4--year COBE--DMR data,
assuming a pure Sachs \&
Wolfe angular spectrum, and with the outputs of BW.

In G96, a (3*961) component signal vector was built
setting together the data of the three maps at 31.5, 53 and 90 GHz.
An analogous (3*961) component noise vector is also built.
Pixels only at galactic longitude above 20$^o$ are taken.
Monopole and dipole components of each map, which are not
physically relevant to the power spectrum estimation,
could be exactly removed by integrating over the first four
components.
We performed a similar analysis using
(2*961) component vectors, based on maps at 53 and 90 GHz
(the 31.5 GHz map, characterized by high noise level, was
not taken into account).
Results are shown
in Fig. \ref{4y}, where we report 1--2--3$\, \sigma$
curves; the top--likelihood point we find is indicated
by an empty triangle, while the G20+* top--likelihood
point (see table \ref{respre}) is indicated by a filled box. 

In BW, a weighted average of the signals at 53 and 90 GHz
is used, as we did in this work. The signal compression
technique is however quite different from the one used
here. The total number of pixels they use is 3890, after
all pixels below the ``custom--cut'', described by Bennet
et al. (1996), are removed. 

The level of consistency between this work and BW can be appreciated
in Fig. \ref{bwl}, where the likelihood distributions on the
spectral index $n$, found by us and BW, are compared. As far as
the distribution on $n$ is concerned, we reproduce even minor features
visible in previous outputs. On the contrary, our $a_{10}$ values tend 
to be slightly smaller and the discrepancy from BW amounts to $\sim 
0.64 \mu K$. 

It may be also interesting to compare our 1--2--3$\, \sigma$
likelihood contours with likelihood contours obtainable from BW.
It ought to be noticed that working out such contours from BW
is far from trivial, as one has to translate their $D'$ and $D''$
parameters into $n$ and $\Omega_\Lambda$ values and this
requires a significant numerical effort. Furthermore, the
peak likelihood curve, at various $n$ values, obtained by
BW, shows features indicating a non--Gaussian behaviour in
respect to $n$. In the $a_{10}$ direction, instead, we have
just the information that 1--$\sigma$ errors (casual) are $\sim 7\, \%$
and only can we assume a Gaussian behaviour.

However, using such information,
it is possible to work out 1--2--3$\, \sigma$ likelihood contours.
In Fig.~\ref{bwc} we compare our and BW curves (thick and
dashed lines, respectively), after displacing BW along the $a_{10} $ axis,
so to have coincident top likelihood points. Those interested
in likelihoods extrapolated from BW results, for the case $\Omega_\Lambda=0$,
$h=0.5$, can work them out from Fig.~\ref{bwc}, shifting the $a_{10}$
axis by 0.64$\, \mu K$. 

The main interest of this figure, however, concerns the actual
distribution of the likelihood, once the non--Gaussian behaviour
is fully taken into account. Fig. 6 confirms that
the distributions along $n$, in this and BW analyses, are pretty similar.
The distributions along $a_{10}$, instead, are different, as is expected.
Substancial discrepancies already exist at 1--$\sigma$ level and they 
are asymmetrical in the two directions of the $a_{10}$ axis. 

\section{PS predictions and a numerical algorithm}

In the preparation of this work, as previously outlined, we
have extensively used the public program CMBFAST.
Besides of the transfer function
$\cal T$$(k)$, this program provides the angular spectra
$C_l$ (for temperature fluctuations and polarization), normalizing
all results and providing the model likelihood on the basis of
BW relations. As previously outlined, this implies
that, for each model, the normalization corresponding to
the best--fit to COBE--DMR 4--year data is selected; the model
likelihood provided by CMBFAST is the one which corresponds to
such best fit.

Making use of the transfer function, a fair deal of large scale
observables can be predicted. For instance, using the Press \&
Schechter (PS) approach, one can evaluate the expected number density
of galaxy clusters. As previously outlined, this is a fair example of
why one needs to go beyond the best--fit normalization. 

Using the fitting formulae worked out in the previous section,
we have therefore built a routine, which may be appended
to CMBFAST, to replace to BW single normalization with possible
normalizations for top likelihood and at the 1--2--3$\, \sigma$
limits (this routine is available on SPOrt web--page,
http://sport.tesre.bo.cnr.it/). Likelihood estimates are also suitably
provided to compare different models, even with different normalizations.

\section{Conclusions}

The main results of this work are: (i) The expression (\ref{piu})
of the 2--dimensional interpolating ``curve'' yielding
the likelihood distribution, for models with given $\Lambda$,
when varying $a_{10}$ and $n$. It is important to stress that,
although the likelihood distribution 
is clearly non--Gaussian, all deviations from a Gaussian
behaviour are fully under control when cubic terms are added.
(ii) The expression (\ref{piu}) is then
implemented by the interpolating expressions of the
$P_i$ coefficients, given by eq.~(\ref{int});
here again, besides of outlining its practical use
in association with Table \ref{param2} parameter values, the reader's
attention is to be attracted to its simplicity. Notice
that, given the values of the 4 parameters $a$, $b$, $c$ and $S$,
our expressions provide the likelihood distribution for any
critical $\Lambda$ model in the $a_{10}$,$n$ plane.
That this likelihood is readily obtainable, for such a wide range of cases,
just assigning a few numerical values, is one of the results of this work.

These results can be completed by an expression allowing
to compare different $\Lambda$ models, in the point
of the $a_{10}$,$\,n$ plane where they reach top likelihood.
In Fig.~\ref{flik} the top likelihood of models is plotted
against $\Omega_\Lambda$. The {\it best--fit} curve shown in
Fig.~\ref{flik} has equation:
\begin{equation} 
{\cal L} = a+b\Omega_\Lambda^S+c\Omega_\Lambda^{2S}
\label{lifit}
\end{equation}
with $a=0.98$, $b=0.34$, $c=-1.24$, $S=1.42$. This expression is
quite similar to eq.~(\ref{int}), yielding the $\Omega_\Lambda$
dependence of the coefficients $P_i$.

As above outlined, the class of models discussed here was previously
treated also by BW, whose results also concerned models with
non--critical density. If we consider only critical $\Lambda$CDM
models, seemingly favoured by current observations, the
likelihood distribution in respect to the spectral index $n$, 
found in BW, is the same as the one found in this work. In respect to
normalization, instead, we find top likelihood for values 
smaller by $\sim 7\, \%$ (see fig.$\, $\ref{bwc}), in respect to BW. 
By itself this is not surprising, owing to various differences between
ours and BW analyses: first of all, the different techniques
adopted to subtract the galactic equatorial band.
Such percentage can be also considered the typical
discrepancy among the amplitudes that different authors obtain
from COBE--DMR data analysis, and the size of the
1--$\sigma$ errorbar for the amplitude. 

Such size of errorbar had already been suggested by BW,
although no justification of their finding is reported in
their article. Besides of a detailed treatment of error
analysis, in this work we provide errors up to 3--$\sigma$,
in the 3--dimensional space spanned by amplitude, spectral
index and $\Lambda$, fully accounting for their deviation
from a Gaussian behaviour.

The above uncertainty of $7\, \%$ makes perhaps redundant
the information on non--Gaussian behaviour at the 1--$\sigma$
level, which implies corrections smaller than $7\, \%$.
However, if a likelihood distribution above 1--$\sigma$
is needed, the results of this work are to be applied;
furthermore, above 1--$\sigma$, deviation from Gaussian behaviour 
approach errorbar size. A typical case when likelihood
distributions above 1--$\sigma$ are needed is when COBE
likelihood is to be considered together with the likelihood
distribution worked out from other experiments.
In this case, the joint likelihood might well reach
its top for a model discrepant more than 1--$\sigma$
from COBE maximum likelihood, but, even if not so,
distributions above 1--$\sigma$ are to be considered
to perform a complete analysis.

As above outlined, our results hold almost independently of
the value of the Hubble parameter $H$. Its impact on $C_l$
up to $l = 30$ is known to be quite limited, and we tested
that, in the range 50--70 km/s/Mpc, we do not need to specify its value.
The same can be said for the baryonic content of the model. Of course,
both $H$ and $\Omega_b$ affect $C_l$ at greater $l$ values.
We have not checked our output against a variation of
the nature of dark matter. If a significant massive neutrino
contribution to dark matter is present (density parameter $\Omega_h$),
some changes of $C_l$ are expected for $l \sim 200$. Such changes
are however rather small (Dodelson et al. 1996), unless neutrinos
have a very late derelativization. The situation might be different
if hot--dark--matter with non--thermal distribution is considered
({\it volatile} models; see Pierpaoli \& Bonometto, 1999).
In any case, however, the impact on $C_l$ of a hot (or 
{\it volatile}) component, down to $l \sim 30$, is expected
to be even smaller than the one arising from $\Omega_b$ shifts.
Hence, our fitting relations can be safely used for
any value of $H$, $\Omega_b$ and for any reasonable
value of $\Omega_h$ originating from massive neutrinos.

\appendix

{{\begin{table}
\label{tabwright}
\begin{tabular}{ccccccccccccccccc}
\tableline
$l$ & $w^{pix}_l$ &  & $l$ & $w^{pix}_l$  & & $l$ & $w^{pix}_l$ & &
$l$ & $w^{pix}_l$ &  & $l$ & $w^{pix}_l$  & & $l$ & $w^{pix}_l$ \\
\tableline
\tableline
 1 & 0.9998 & &  6 & 0.9966 & & 11 & 0.9893 & & 16 & 0.9780 & & 21 &
0.9629 & & 26 & 0.9439 \\ 
 2 & 0.9995 & &  7 & 0.9954 & & 12 & 0.9873 & & 17 & 0.9753 & & 22 &
0.9594 & & 27 & 0.9397 \\ 
 3 & 0.9990 & &  8 & 0.9941 & & 13 & 0.9853 & & 18 & 0.9724 & & 23 &
0.9557 & & 28 & 0.9353 \\ 
 4 & 0.9984 & &  9 & 0.9927 & & 14 & 0.9830 & & 19 & 0.9694 & & 24 &
0.9519 & & 29 & 0.9308 \\ 
 5 & 0.9976 & & 10 & 0.9911 & & 15 & 0.9806 & & 20 & 0.9662 & & 25 &
0.9480 & & 30 & 0.9262 \\
\tableline
\end{tabular}
\caption{COBE pixalization function values; see text.}
\end{table}}

\section{Appendix A}
The sky temperature measured in a direction
$\hat n$$\equiv \theta,\,\phi$ is a convolution of the 
temperatures around ($\hat n$) with the beam pattern.
Temperatures are then averaged inside each pixel area.
Let then $w^{pix}(\theta,\,\phi)$ be the filter function,
that models the pixel shape.
Assuming it to be symmetric around the polar axis (that
we choose in the $\theta$ direction), $w^{pix}$ depends only 
on the angular coordinate $\theta$ and can be expanded in
Legendre polynomials; the coefficients of the expansion then read:
\begin{equation}
w^{pix}_l={\int_0^{\pi}d \theta \sin \theta w^{pix}(\theta) 
P_l(cos\theta)\over \int_0^{\pi}d \theta \sin\theta
 w^{pix}(\theta)}\,.
\label{aa1}
\end{equation}
For COBE pixelization $w^{pix}$ is a circular top--hat
window with angular area $\Omega_{pix}=4\pi/6144$; let then ${\bar
\theta}$ be such that
\begin{equation}
\int_0^{2\pi}d\phi\int_0^{{\bar \theta}}d\theta=\Omega_{pix}\,,
\end{equation}
then the eq~(\ref{aa1}) becomes 
\begin{equation}
w^{pix}_l=-{2\pi \over \Omega_{pix}}\int _1^{cos\bar{\theta}}dx~ P_l(x)=
-{2\pi \over \Omega_{pix}}\int _1^{cos\bar{\theta}}dx~{1\over 2^l l!}
{d^l(x^2-1)^l\over dx^l}= 
$$$$
-{2\pi \over \Omega_{pix}}{1\over 2^l l!}\int _1^{cos\bar{\theta}}dx{d\over dx}
\Bigg[{d^{l-1}(x^2-1)^l\over dx^{l-1}}\Bigg]=
-{2\pi \over \Omega_{pix}}{1\over 2^l l!}\Bigg[{d^{l-1}(x^2-1)^l\over 
dx^{l-1}}\Bigg]_1^{cos\bar{\theta}}.
\end{equation}
The values of $w^{pix}_l$ are reported in the table below.

\section{Appendix B}
By integrating the likelihood $\cal L$
over the first 4 components of the vector ${\bf c}$, we can remove the
contribution of the monopole and dipole terms. Such integration can
be easily performed in the case of Gaussian distributions. 

In fact, let us rewrite the expression of
\begin{equation}
{\cal L}(c_{\mu}) = {1 \over \sqrt{(2\pi)^{N} det{\bf M}}}
\exp[-(1/2){\bf c}^{\rm T}{\bf M}^{-1}{\bf c}] ~,
\label{ab1}
\end{equation}
using a greek index for the components from 1 to 4, and $x$ or
$y$ for the components from 5 to 961. The integral of 
eq. (\ref{ab1}) reads then
\begin{equation}
\int dc_{\mu}~{\cal L}(c_{\mu})=
{1\over \sqrt{(2 \pi)^{N}det{\bf M}}}
\int dc_{\mu}~\exp~(-{1\over 2}c_{\mu} {M}^{-1}_{\mu \nu} c_{\nu}-
{1 \over 2}c_x {M}^{-1}_{xy} c_y- c_{\mu} {M}^{-1}_{\mu x} c_x).
\end{equation}
Let be $J_{\mu} \equiv {M}^{-1}_{\mu x} c_x$, then
\begin{equation}
\int dc_{\mu}~{\cal L}(c_{\mu})=
{1\over \sqrt{(2 \pi)^{N}det{\bf M}}}\exp(-{1 \over 2}
c_x {M}^{-1}_{xy} c_y)\times
$$$$
\times\int dc_{\mu}~\exp~(-{1\over 2}c_{\mu} {M}^{-1}_{\mu \nu} c_{\nu}-
2c_{\mu} J_{\mu})\,.
\end{equation}
Let then be $\tilde{c_{\mu}}=c_{\mu}+b_{\mu \sigma}$,
where $ b_{\mu \sigma}$ is a generic matrix $(4 \times 4)$ independent
from $c_{\mu}$; using $\tilde c_\mu$ as integration variables, 
the exponent in the integration becomes
\begin{equation}
c_{\mu} {M}^{-1}_{\mu \nu} c_{\nu}+ 2c_{\mu} J_{\mu}=
(\tilde{c_{\mu}}- b_{\mu \sigma}) J_{\sigma} {M}^{-1}_{\mu \nu}
(\tilde{ c_{\nu}}- b_{\nu \tau} J_{\tau})+
$$$$
+2 (\tilde{c_{\mu}}- b_{\mu \sigma}) J_{\mu}=
\tilde{c_{\mu}}{M}^{-1}_{\mu \nu}\tilde{ c_{\nu}}- b_{\mu \sigma}
J_{\sigma} {M}^{-1}_{\mu \nu}\tilde{ c_{\nu}}+
$$$$
-\tilde{c_{\mu}}
{M}^{-1}_{\mu \nu} b_{\nu \tau} J_{\tau}+ b_{\mu \sigma} J_{\sigma}
{M}^{-1}_{\mu \nu} b_{\nu \tau} J_{\tau}+2 \tilde{c_{\mu}} J_{\mu}-
2 J_{\mu} b_{\mu \sigma} J_{\sigma}~.
\label{ab2}
\end{equation}
If ${\bf b}$ is the inverse of the $(4 \times 4)$ matrix
$({M}^{-1})_{\mu,\nu}$, the eq. (\ref{ab2}) simplifies into
\begin{equation}
c_{\mu} {M}^{-1}_{\mu \nu} c_{\nu}+ 2c_{\mu} J_{\mu}=
\tilde{c_{\mu}}{M}^{-1}_{\mu \nu}\tilde{ c_{\nu}}-J_{\nu} 
\tilde{ c_{\nu}}- J_{\mu}\tilde{ c_{\mu}}+ J_{\mu}b_{\mu \sigma}
J_{\sigma}+
$$$$
+2\tilde{c_{\mu}} J_{\mu}-2 J_{\mu}b_{\mu \sigma}
J_{\sigma}=\tilde{c_{\mu}}{M}^{-1}_{\mu \nu}\tilde{ c_{\nu}}-
J_{\mu}b_{\mu \sigma} J_{\sigma}~.
\end{equation}
Henceforth, performing the integration, we have
\begin{equation}
\int dc_{\mu}~{\cal L}={1\over \sqrt{(2 \pi)^{N}det{\bf M}}}
\exp[-{1 \over 2}(c_x {M}^{-1}_{xy} c_y+J_{\mu}b_{\mu \sigma} 
J_{\sigma})]\int dc_{\mu}\exp(-{1 \over 2} \tilde{c_{\mu}}
{M}^{-1}_{\mu \nu}\tilde{ c_{\nu}})=
$$$$
={1\over \sqrt{(2 \pi)^{N}det{\bf M}}}(2 \pi)^2
{1\over \sqrt{(det{\bf b}^{-1})}}\exp[-{1 \over 2}(c_x {M}^{-1}_{xy} c_y+
c_x {M}^{-1}_{x \mu}b_{\mu \sigma} {M}^{-1}_{\sigma y} c_y)]=
$$$$
={1\over \sqrt{(2 \pi)^{N-4}det {\bf {\tilde M}}}} \exp[-{1 \over 2}
(c_x \tilde{M}^{-1}_{xy} c_y)]~;
\end{equation}
the new covariance matrix ${\bf {\tilde M}}$ has $((N-4) \times
(N-4))$ dimension.  

Hence, the integration of the first four components of the vector
${\bf c}$ is equivalent to replace the matrix ${\bf M}$
with ${ {\tilde M}_{xy}}={M}^{-1}_{xy}+{M}^{-1}_{x \mu}b_{\mu \sigma}
{M}^{-1}_{\sigma y}$ (and $N$ with $N-4$)  in eq~(\ref{ab1}).

{\begin{figure*}[htb]
\centerline{\epsfxsize=15truecm
\epsfbox{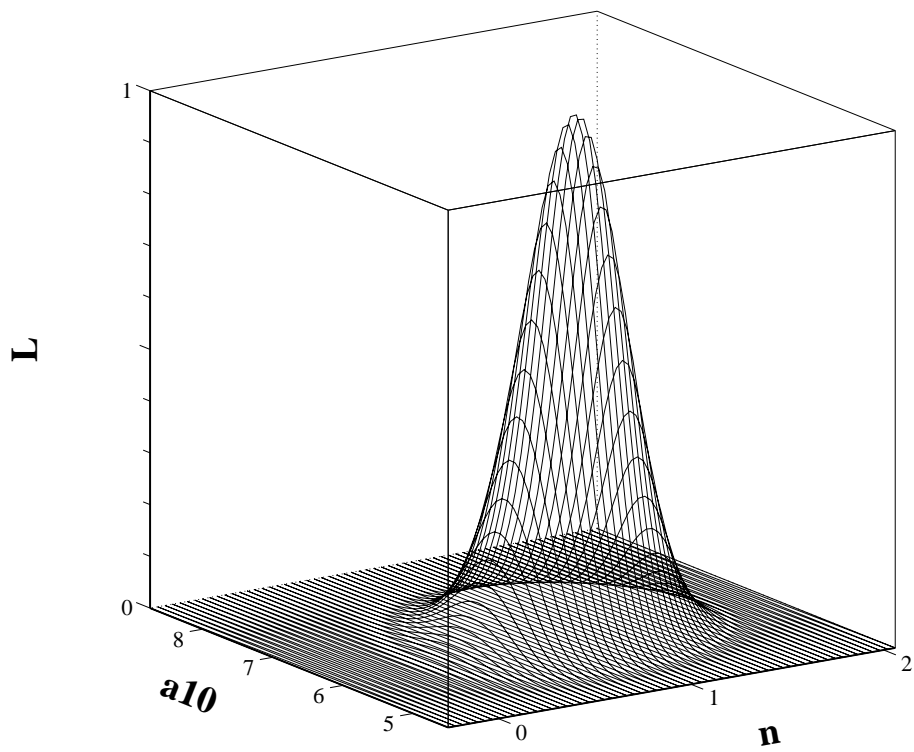}}
\caption{Likelihood function for $\Omega_{\Lambda}=0$ and
$h=0.5$; $a_{10}$ is expressed in $\mu K$. \label{3dlike}}
\label{3dlike}
\end{figure*}}

{\begin{figure*}[htb]
\centerline{\epsfxsize=7truecm
\epsfbox{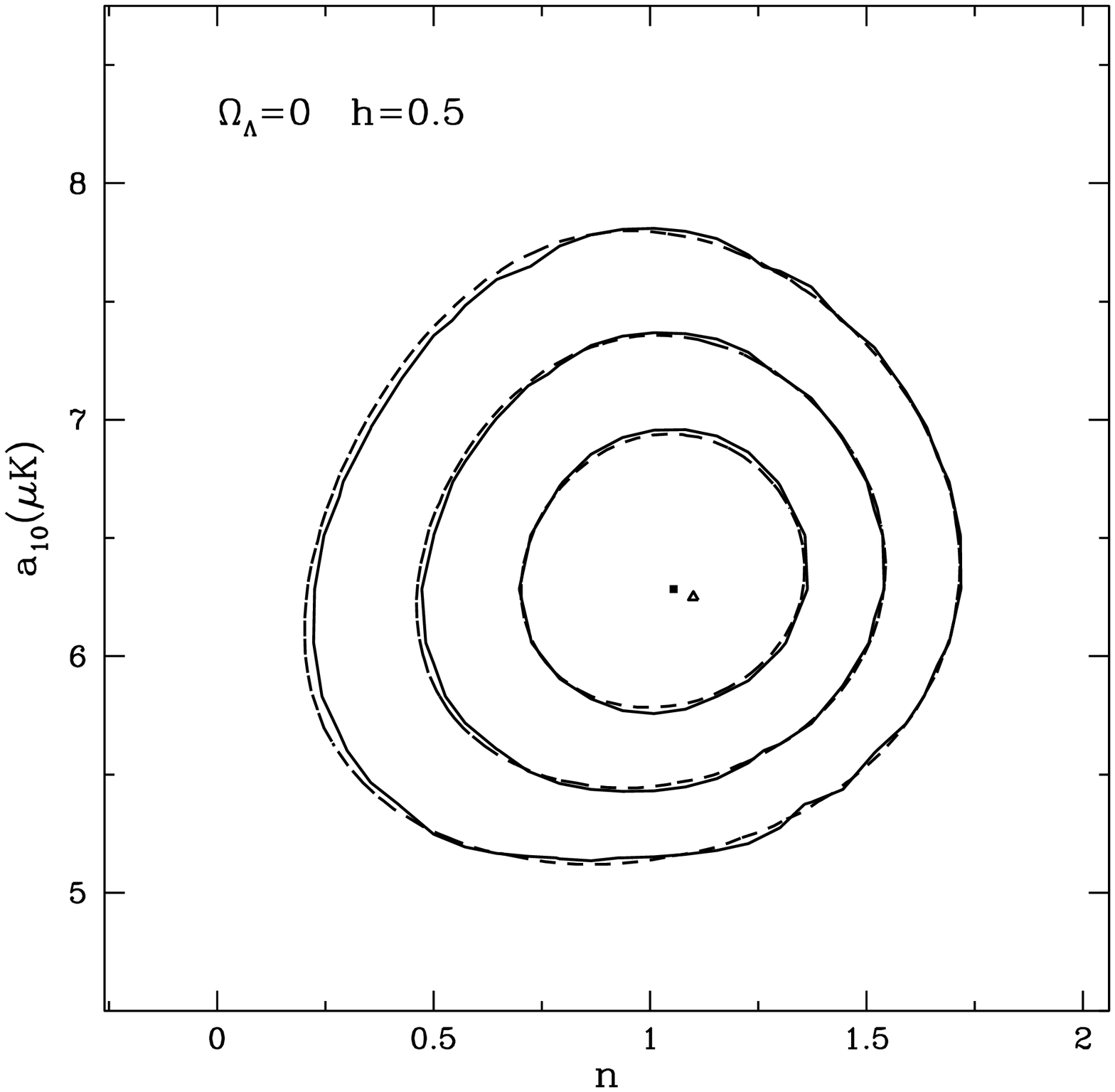}}
\centerline{\epsfxsize=7truecm
\epsfbox{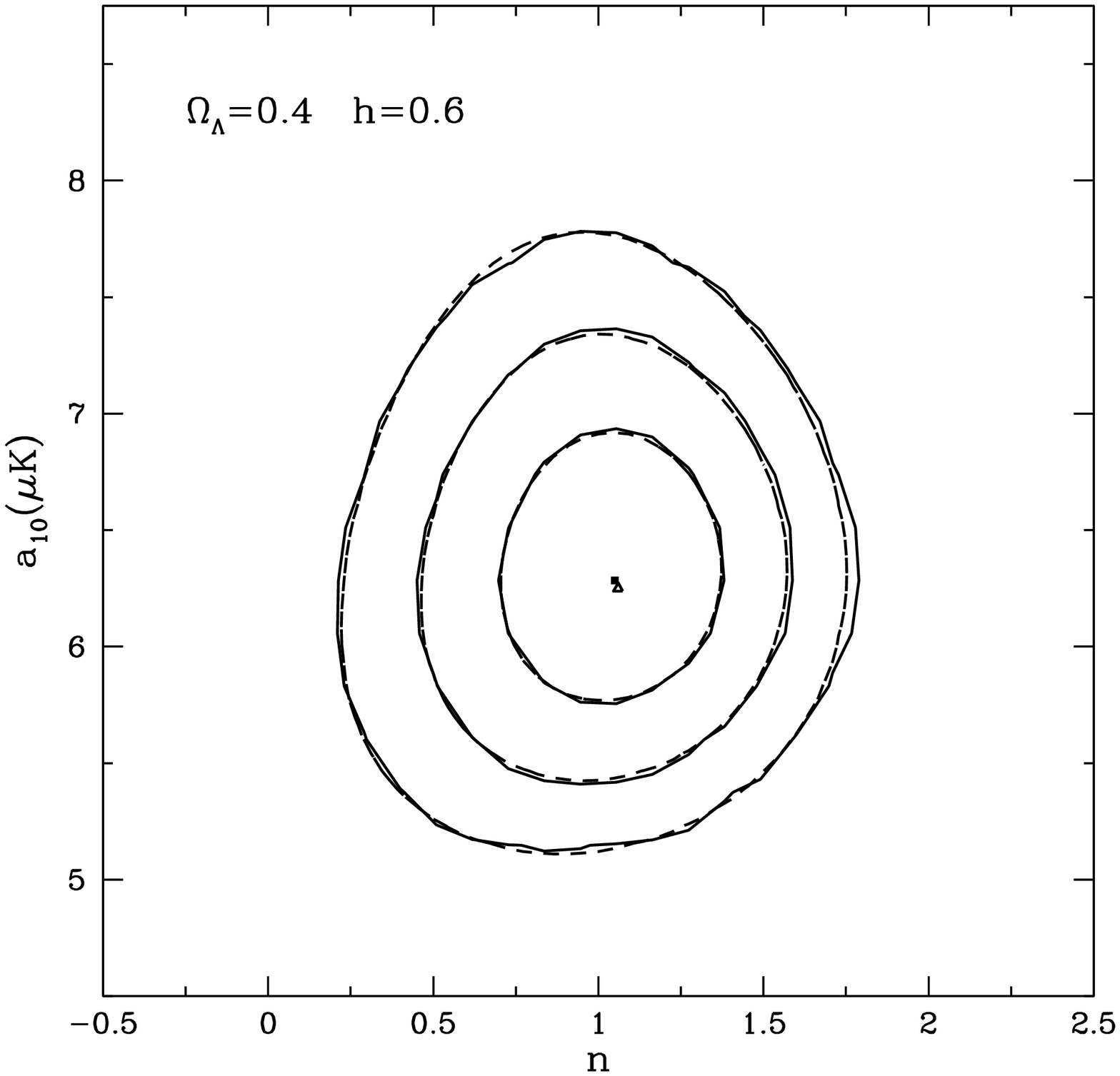}}
\centerline{\epsfxsize=7truecm
\epsfbox{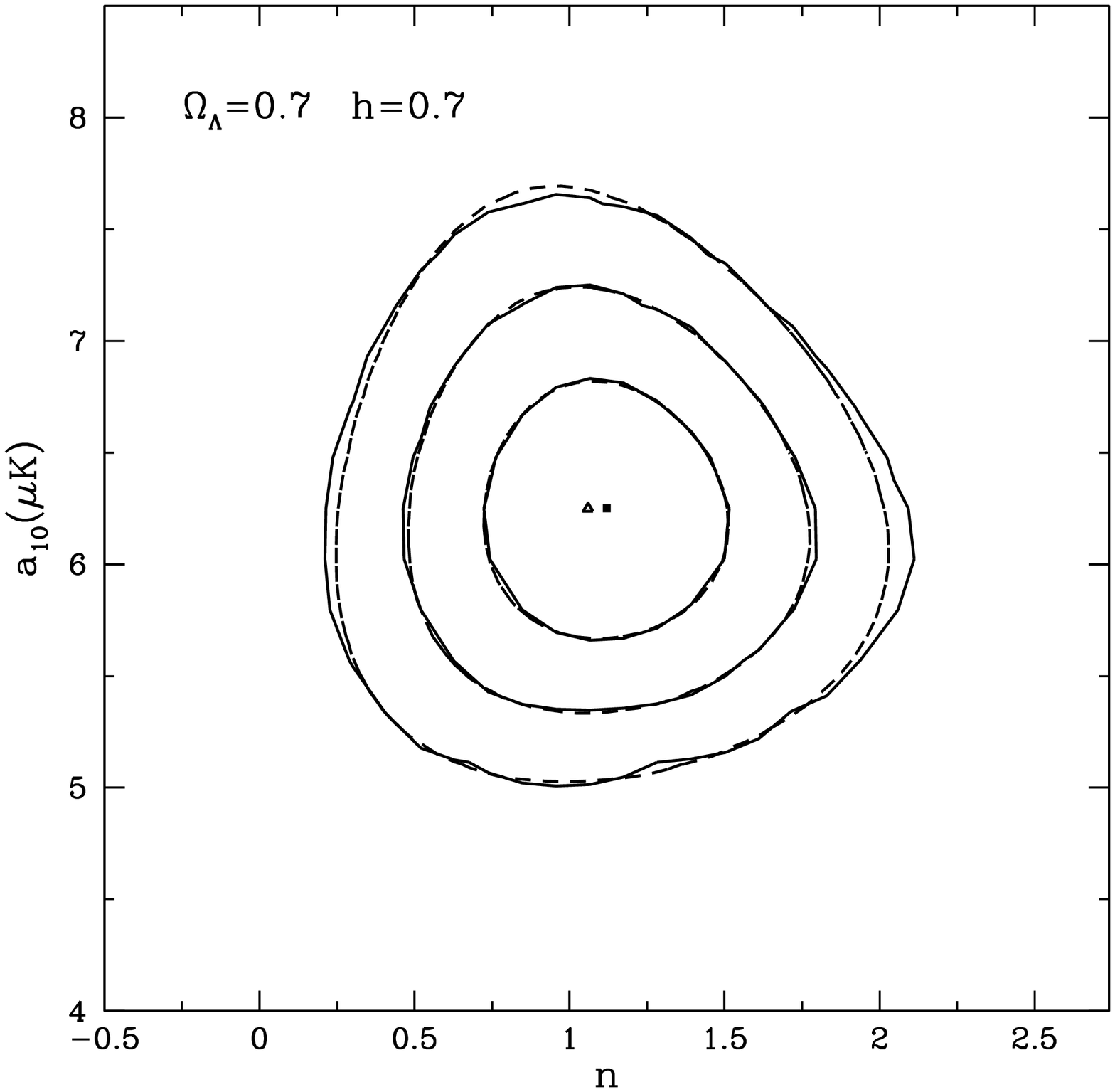}}
\caption{Examples of comparison
between the 1--2--3$\, \sigma$ confidence levels and likelihood
peaks directly obtained from the likelihood function
(continuous lines, filled box) and the contours
obtained using the fitting formula, eq. (\ref{piu}) (dashed
lines, empty triangles).} 
\label{likeh}
\end{figure*}} 

{\begin{figure*}[htb]
\centerline{\epsfxsize=18truecm
\epsfbox{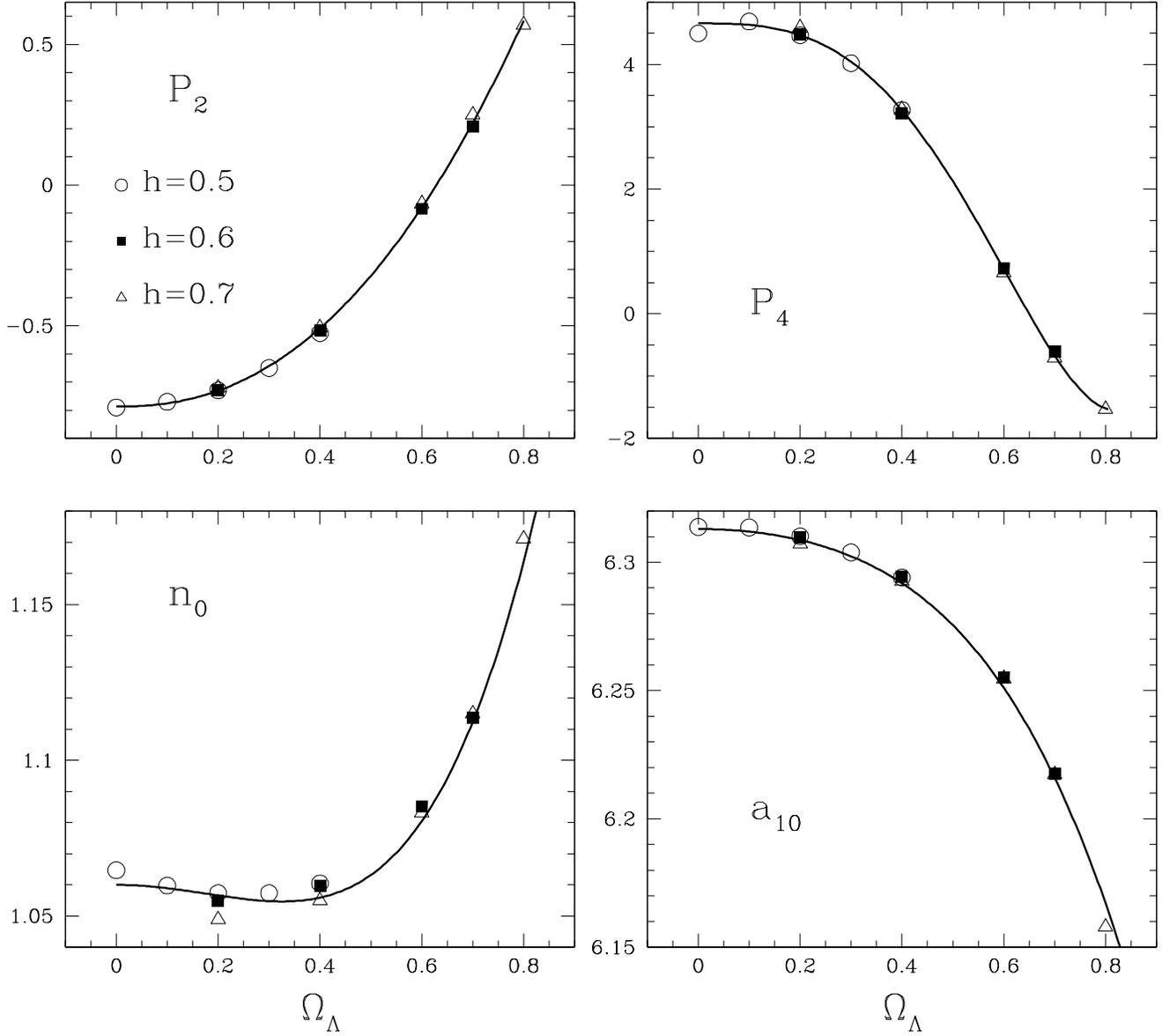}}
\figcaption{Examples of the fits between
$n^0$, $a_{10}^o$ and the parameters $P_i$ and the analytic expression
\ref{int}. Different symbols are related to the values of
$h$. Discrepancies between parameters and fitting expression are 
$< 1\%$.}
\label{p2p4}
\end{figure*}}

{\begin{figure*}[htb]
\centerline{\epsfxsize=18truecm
\epsfbox{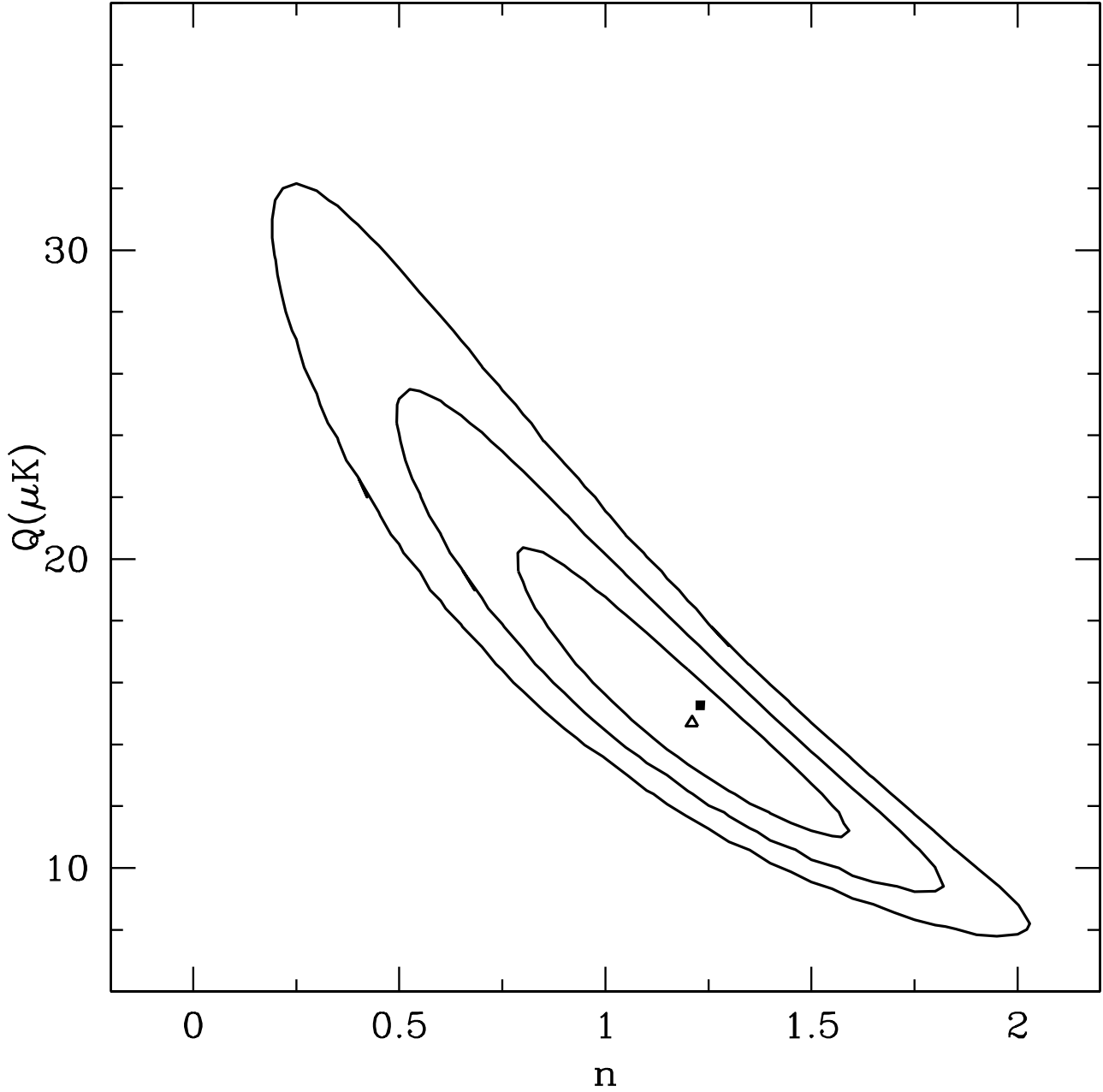}}
\figcaption{The 1--2--3$\,\sigma$ confidence levels on
$Q$--$n$ plane from 4--years COBE data, using a Sachs \& Wolfe
spectrum. The empty triangle is the top likelihood point we obtain.
For the sake of comparison, the filled box is the top likelihood
point obtained in G20+* (see text).
\label{4y}}
\end{figure*}}

{\begin{figure*}[htb]
\centerline{\epsfxsize=18truecm
\epsfbox{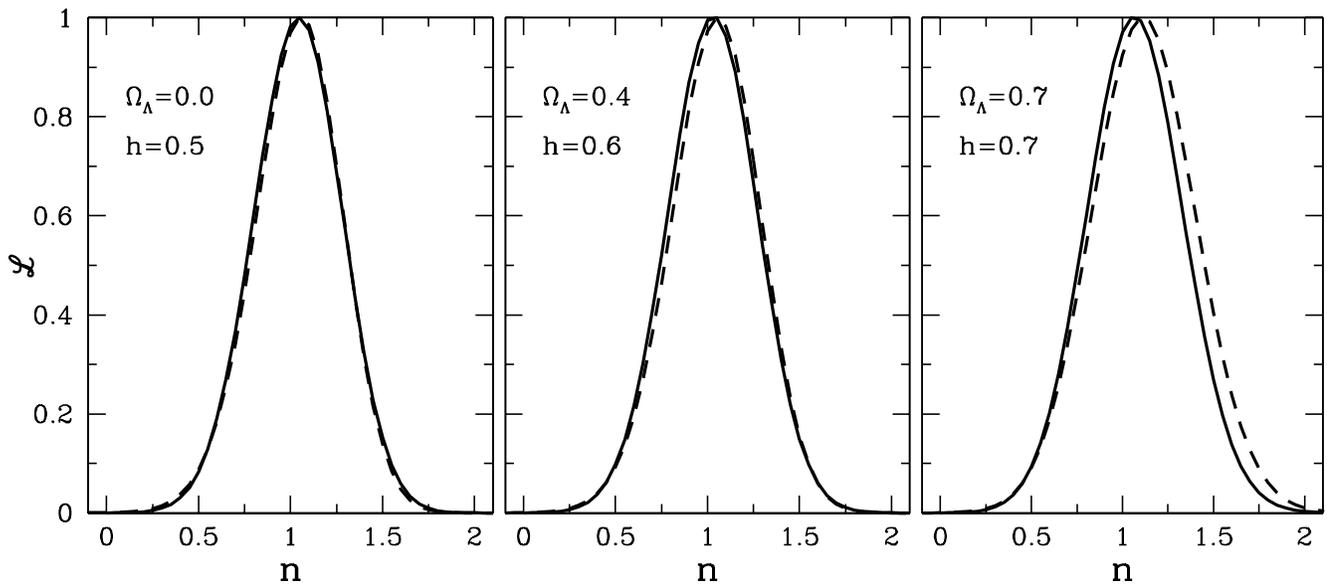}}
\caption{Likelihood
distribution against $n$ from BW (dashed line) and our work (continuous
line). }
\label{bwl}
\end{figure*}}

{\begin{figure*}[htb]
\centerline{\epsfxsize=18truecm
\epsfbox{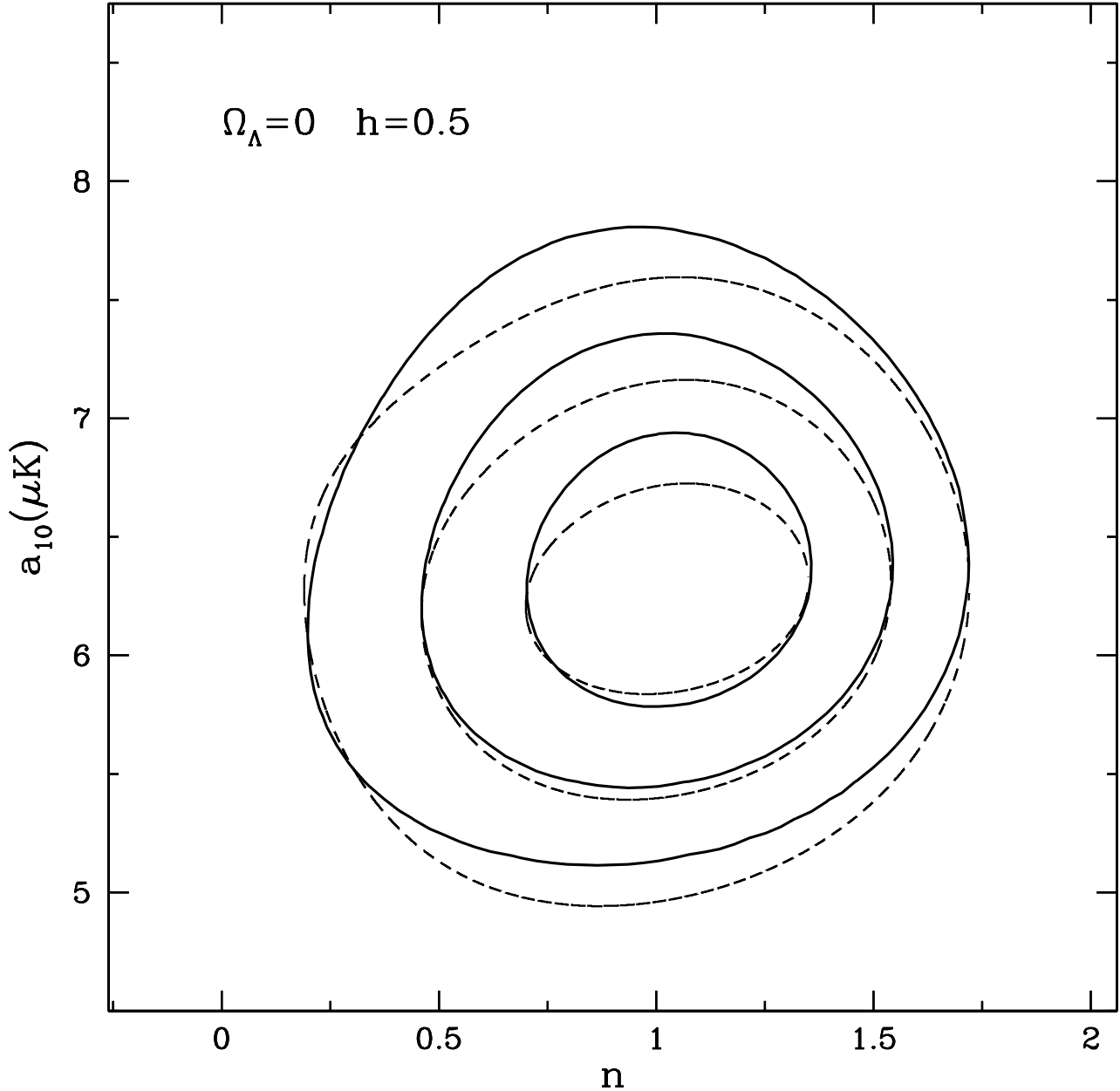}}
\caption{Confidence contours are
shown for a model with $\Omega_{\Lambda}=0.$ and $h=0.5$
(thick lines) and compared with confidence contours obtainable from BW
(dashed lines), displaced to overlap top likelihood points. This
figure exhibits the effect of taking suitably into account the deviations
from a Gaussian behaviour. }
\label{bwc}
\end{figure*}}

{\begin{figure*}[htb]
\centerline{\epsfxsize=18truecm
\epsfbox{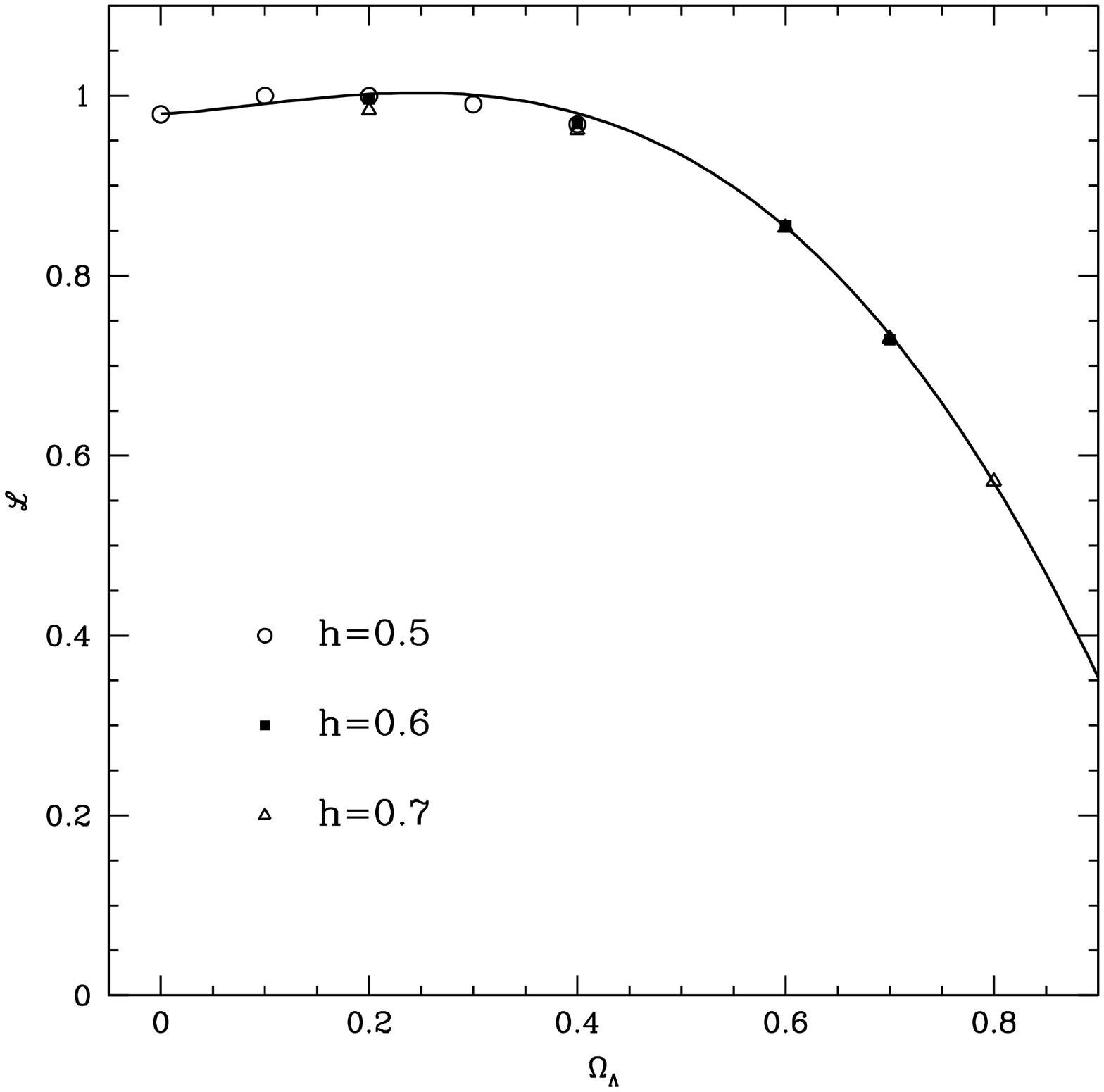}}
\caption{Top values of the likelihood for models with
different $\Omega_{\Lambda}$. The values of $a_{10}$ and $n$ giving
such likelihood are obtained from expression \ref{lifit}.} 
\label{flik}
\end{figure*}}


\begin{thebibliography}{}

\bibitem{beeal94} Bennett, C.L., et al. 1994, ApJ, 436, 423

\bibitem{beeal96} Bennett, C.L., et al. 1996, ApJ, 646, L1

\bibitem{be} Bond, J.R. \&\ Efstathiou, G., 1987, MNRAS, 226, 665

\bibitem{bond} Bond, J.R., 1995, Phys. Rev. Lett., 74, 4369

\bibitem{bp} Bonometto, S.A. \&\ Pierpaoli, E., 1998, NewA., 3, 391

\bibitem{busu} Bunn, E.F. \&\ Sugiyama, N., 1995, ApJ, 446, 49

\bibitem{buwh} Bunn, E.F. \&\ White, M., 1997, ApJ, 480, 6 (BW)

\bibitem{cfmg} Contaldi, C.R., Ferreira, P.G., Magurijo, J. \&\
G\'orski, K.M., 2000, ApJ, 534, 25

\bibitem{boom} de Bernardis, P., et al. 2000, Nature, 404, 995

\bibitem{dgs} Dodelson, S., Gates, E. \&\ Stebbins, A., 1996, ApJ,
467, 10

\bibitem{eke1} Eke, V.R., Cole, S. \&\ Frenk, C.S., 1996, MNRAS, 282, 263  

\bibitem{eke2} Eke, V.R., Cole, S., Frenk, C.S., \&\ Henry, J.P., 1998,
MNRAS, 298, 1145

\bibitem{fmg} Ferreira, P.G., Magurijo, J. \&\ G\'orski, K.M., 1998,
ApJ, 503, L1

\bibitem{gira} Girardi, M., Borgani, S., Giuricin, G., Mardirossian, F., 
\&\ Mezzetti, M., 1998, ApJ, 506, 45

\bibitem{g94} G\'orski, K.M., 1994, ApJ, 430, L85 (G94)

\bibitem{goeal94} G\'orski, K.M., Hinshaw, G., Banday, A.J., Bennett,
C.L., Wright, E.L., Kogut, A., Smoot, G.F. \&\ Lubin, P.M., 1994, ApJ,
430, L89

\bibitem{goeal96} G\'orski, K.M., Banday, A.J., Bennett, C.L.,
Hinshaw, G., Kogut, A., Smoot, G.F., \&\ Wright, E.L., 1996, ApJ, 464,
L11 (G96)

\bibitem{g97} G\'orski, K.M., "Microwave Background Anisotropies",
proceedings of the XVIth Moriond Astrophysics Meeting, Editions
Frontieres, p.77; astro-ph/9701191

\bibitem{maxima} Hanany, S., et al., 2000, astro-ph/0005123

\bibitem{hinsh} Hinshaw, G., Banday, A.J., Bennett, C.L., G\'orski,
K.M., Kogut, A., Smoot, G.F., \&\ Wright, E.L., 1996, ApJ, 464, L17

\bibitem{mnv} Muciaccia, P.F., Natoli, P. \&\ Vittorio, N., 1997, ApJ,
488, L63

\bibitem{nfs} Novikov, D., Feldman, H. \&\ Shandarin, S., 1998,
astro-ph/9809238 

\bibitem{pvf} Pando, J., Valls-Gabaud \&\ Fang, L.Z., 1998,
Phys. Rev. Lett., 81, 4568

\bibitem{perl} Perlmutter, S., et al. 1999, ApJ, 517, 565

\bibitem{pb} Pierpaoli, E. \&\ Bonometto, S.A., 1999, MNRAS, 305, 425

\bibitem{riess} Riess, A.G., et al. 1998, AJ, 116, 1009

\bibitem{sw} Sachs, R.K. \&\ Wolfe, A.M., 1967, ApJ, 147, 73

\bibitem{sz} Seljak, U. \&\ Zaldarriaga, M., 1996, ApJ, 469, 437

\bibitem{smoeal} Smoot, G.F., et al. 1992, ApJ, 396, L1

\bibitem{sgba} Stompor R., \&\ G\'orski, 1994, ApJ 422, L41

\bibitem{sgba} Stompor, R., G\'orski, K.M. \&\ Banday, A.J., 1995a,
MNRAS, 277, 1225

\bibitem{sgbb} Stompor, R., G\'orski, K.M. \&\ Banday, A.J., 1995b,
astro-ph/9502035

\bibitem{tegbu} Tegmark, M. \&\ Bunn, E.F., 1995, ApJ, 455, 1

\bibitem{teg} Tegmark, M., 1997, Phys. Rev. D55, 5895; astro-ph/9611174

\bibitem{tegha} Tegmark, M. \&\ Hamilton, A., 1997, astro-ph/9702019


\bibitem{wreal94a} Wright, E.L., Smoot, G.F., Kogut, A., Hinshaw, G.,
Tenorio, L., Lineweaver, C., Bennett, C.L. \&\ Lubin, P.M., 1994, 420, 1

\bibitem{wreal94b} Wright, E.L., Smoot, G.F., Bennett, C.L. \&\ Lubin,
P.M., 1994, 436, 443 

\bibitem{wreal96} Wright, E.L., Bennett, C.L., G\'orski, K.M.,
Hinshaw, G. \&\ Smoot, G.F., 1996, ApJ, 464, L21

\end{thebibliography}
\end{document}